# Do we measure novelty when we analyze unusual combinations of cited references? A validation study of bibliometric novelty indicators based on F1000Prime data

Lutz Bornmann[*], Alexander Tekles[*+], Helena H. Zhang[**§], Fred Y. Ye[**$]

*Division for Science and Innovation Studies
Administrative Headquarters of the Max Planck Society
Hofgartenstr. 8,
80539 Munich, Germany.
Email: bornmann@gv.mpg.de
Email: alexander.tekles.extern@gv.mpg.de

[+]Ludwig-Maximilians-Universität Munich
Department of Sociology
Konradstr. 6
80801 Munich, Germany.

** Jiangsu Key Laboratory of Data Engineering and Knowledge Service
School of Information Management
Nanjing University
Nanjing 210023, China

and

International Joint Informatics Laboratory (IJIL)
Nanjing University – University of Illinois,
Nanjing – Champaign,
China – USA

§ Email: 137356390@qq.com
$ Email: yye@nju.edu.cn

**Abstract**

Lee, Walsh, and Wang (2015) – based on Uzzi, Mukherjee, Stringer, and Jones (2013) – and Wang, Veugelers, and Stephan (2017) proposed scores based on cited references (cited journals) data which can be used to measure the novelty of papers (named as novelty scores U and W in this study). Although previous research has used novelty scores in various empirical analyses, no study has been published up to now – to the best of our knowledge – which quantitatively tested the convergent validity of novelty scores: do these scores measure what they propose to measure? Using novelty assessments by faculty members (FMs) at F1000Prime for comparison, we tested the convergent validity of the two novelty scores (U and W). FMs' assessments not only refer to the quality of biomedical papers, but also to their characteristics (by assigning certain tags to the papers): for example, are the presented findings or formulated hypotheses novel (tags "new findings" and "hypothesis")? We used these and other tags to investigate the convergent validity of both novelty scores. Our study reveals different results for the novelty scores: the results for novelty score U are mostly in agreement with previously formulated expectations. We found, for instance, that for a standard deviation (one unit) increase in novelty score U, the expected number of assignments of the "new finding" tag increase by 7.47%. The results for novelty score W, however, do not reflect convergent validity with the FMs' assessments: only the results for some tags are in agreement with the expectations. Thus, we propose – based on our results – the use of novelty score U for measuring novelty quantitatively, but question the use of novelty score W.

# 1 Introduction

For many years, peer review was the only method used for research assessments (Bornmann, 2011). Since the 1980s, however, peer review has been increasingly complemented by bibliometric indicators. According to Belter (2018), "bibliometrics are now regularly used in national research evaluation exercises in countries around the world … and are incorporated in many high profile university rankings such as the Times Higher Education ranking [see https://www.timeshighereducation.com/world-university-rankings] and the Shanghai Ranking [see http://www.shanghairanking.com]. Bibliometrics are also routinely used at the individual level to evaluate researchers for promotion and tenure and, in some cases, to make funding decisions." Today, bibliometrics is often employed in so-called informed peer review processes, in which peers assess scientific units (e.g., institutions) based on bibliometric indicators (and further metrics) (Gunashekar, Wooding, & Guthrie, 2017). Berger and Baker (2014) list some factors which drive the increasing use of bibliometrics in research evaluation: "shrinking budgets, heightened competition, increasing scientific research output, and longevity of the major metrics. Also important are what may seem to be contradictory trends; technology and globalization. Technology has spurred the enhancement of tools used to generate bibliometric data, but also has provided more opportunities to measure different data, causing some in the research community to question the importance of the standard metrics" (p. 83). Today, research assessments based on bibliometrics are integral parts of many scientific activities (Moed, 2017).

Indicators are measurable quantities which are "substitutes for something less readily measurable and is presumed to associate with it without directly measuring it" (Wilsdon et al., 2015, p. 5). Bornmann and Marewski (2019) coined the term bibliometrics-based heuristics for scientific performance assessments if the decisions are based on only two pieces of information from the available indicators space: citations and publications. Most of the indicators which have been developed in bibliometrics hitherto focus on four major areas: "productivity, collaboration, research topics, and citation impact" (Belter, 2018). For example, citation counts are frequently used proxies for scientific impact, although past research might influence current research which is not reflected in citations (e.g., because the impact is induced by inspiring conference presentations).

In recent years, another area of indicators has been added to the established areas by introducing indicators which might capture the novelty of research described in publications (Uzzi et al., 2013; Wang et al., 2017). Novel combinations of existing knowledge are seen as prerequisites for important research able to inspire later research. The novelty indicators are usually based on counting unusual combinations of cited references in papers. Since novelty is seen as related to unusual combinations of existing knowledge, this concept has been transferred to bibliometrics based on cited references (as proxies of existing knowledge in the archives). The new area of novelty indicators is of specific interest for the field of research evaluation, since this is interested in research leading to new breakthroughs (Winnink, Tijssen, & van Raan, 2016): "Scientific progress is driven by important, infrequent discoveries that cannot be readily identified and quantified, which makes research assessment very difficult" (Rodríguez-Navarro, 2016).

When new indicators are introduced in scientometrics, the first task is to measure whether these indicators really measure what they propose to measure (Panel for Review of Best Practices in Assessment of Research et al., 2012). In a recent empirical study, Tahamtan and Bornmann (2018) interviewed authors of landmark papers in bibliometrics to investigate whether the great potential of their papers is reflected in the cited references. Tahamtan and Bornmann (2018) conclude that the ideas for the papers might not necessarily have been inspired by past publications. The cited literature "seems to be important for the contextualization of the idea in the field of scientometrics. Instead, we found that … [the] ideas are the result of finding solutions to practical problems, result from discussions with colleagues, and profit from interdisciplinary exchange" (p. 906). Thus, the results of Tahamtan and Bornmann (2018) question the common use of cited references to measure novelty. In other words, the results of this qualitative study might be a first hint that cited references are not useful data for measuring the novelty of papers.

Thus, we started to conduct this additional quantitative study in which we explore the concepts of measuring novelty further. We compared novelty scores with novelty assessments by peers (experts in the corresponding fields) to test the convergent validity of the proposed metrics. The convergent validity of the scores would be given, if they were in fact positively related to the assessment by peers. The study of convergent validity based on quantitative empirical data seems necessary, since Lee et al. (2015) already "encourage use of such metrics as a new tool for science policy indicators" (p. 694).

## 2 Measuring creativity based on novelty (scores)

In the last two decades, scholars have used various types of bibliographic information and methods to identify breakthrough-class papers, such as Redner (2005), Schneider and Costas (2017), and Ponomarev, Williams, Hackett, Schnell, and Haak (2014). Winnink et al. (2016) investigated the specific characteristics of breakthrough-class papers which they identified using various bibliographic information (e.g., cited references or number of authors) and methods proposed hitherto. Identifying and measuring novelty in science based on quantitative indicators has a long tradition (in scientometrics and beyond), since novelty is regarded as one of the most important conditions for excellent research.

Creativity – the medium in which novel directions in research emerge – has frequently been interpreted as an evolutionary search process within structured spaces (i.e., disciplinary archives) including certain elements (i.e., single knowledge proposals). The process of imaging novel combinations of elements from this space – crossing established (disciplinary) boundaries – is interpreted as creativity (Burt, 2004; Lee et al., 2015; Mednick, 1962;

Simonton, 2003). In other words, recombination processes are seen as the main mechanisms for being creative which manifest in publishing breakthrough-class papers (Kaplan & Vakili, 2015). Koestler (1964) introduced the term "bisocitation" to describe these processes which connect well-known, but previously unconnected entities. There is an emerging line of research which has operationalized this approach by measuring the novelty of journals, patents, and research proposals (e.g., Boudreau, Guinan, Lakhani, & Riedl, 2016; Fleming, 2001; Packalen & Bhattacharya, 2017; Verhoeven, Bakker, & Veugelers, 2016). Basically, all these studies follow "the combinatorial novelty literature, which views research as a problem solving process involving various combinatorial aspects so that novelty comes from making unusual combinations of preexisting components" (Wang, Lee, & Walsh, 2018, p. 1074). These combinatorial processes are seen as creative acts.

Uddin and Khan (2016) used the combination of usual and unusual keywords by authors to identify knowledge innovation and the introduction of new concepts. A similar measurement of "thematic novelty" of papers has been used by Carayol, Llopis, and Lahatte (2016) which is based on the rareness of keywords combinations. These authors found, for instance, that high thematic novelty of papers is associated with high citation impact. Schubert and Schubert (1997) and Schubert (2013) analyzed new combinations of terms in the title of papers published in chemistry. They used the method to detect emergent areas in chemistry. Warman, Bradford, and Moles (2013) based their study on the novelty of MeSH term combinations. MeSH terms thematically classify papers which are published in the field of biomedicine. The authors compared the MeSH term combinations related to grant proposals with existing combinations in the literature. They found, for instance, that "evaluators systematically give lower scores to research proposals that are closer to their own areas of expertise and to those that are highly novel. The patterns are consistent with biases associated with boundedly rational evaluation of new ideas" (p. 2765). Foster, Rzhetsky, and Evans (2015) used combinations of chemicals reported in the literature to measure novelty and found that high novelty is related to high citation impact.

In this study, combining with F1000Prime qualitative recommendations, we investigated two quantitative approaches which used unusual combinations of cited literature in papers as measures of novelty: (1) Uzzi et al. (2013) – revised by Lee et al. (2015) – and (2) Wang et al. (2017). Both approaches are based on the premise that creative ideas in science are based on novelty which is reflected in unusual combinations of cited literature (Uzzi et al., 2013). In other words, many unusual combinations point to creative research for which the chance of becoming highly-cited or breakthrough-class status is high.

## 3 Methods

### 3.1 Post-publication peer review at F1000Prime

F1000Prime is a post-publication peer review system in the biomedical area. According to Tennant et al. (2018), it is a recommendation service which provides "post-publication evaluation and recommendation of significant articles, often through a peer-nominated consortium". So-called Faculty Members (FMs) recommend, score, and classify papers for inclusion in the F1000Prime database. To ensure that the literature in the relevant topic areas has been adequately covered in the database, the tables of contents of major general and specialist journals in biomedicine are scanned. F1000Prime recommends on average more than 1,000 papers each month, corresponding to a fraction of all published papers in biology and the medical sciences. Thus, the majority of papers published in these areas are not covered in the F1000Prime database (see below). When FMs recommend a paper for F1000Prime, they also write a brief review explaining what makes the paper so important and rate it as "good", "very good" or "exceptional" (equivalent to scores of 1, 2 or 3 stars, respectively). F1000Prime uses the individual scores to calculate the total scores for each paper (FMs recommendations), which are used to rank the papers in each discipline.

FMs also assess the papers with the following tags, if appropriate:[1]

- Confirmation: article validates previously published data or hypotheses
- *Controversial*: article challenges established dogma
- Good for teaching: key article in a field and/or is well written
- **Hypothesis**: article presents an interesting hypothesis
- Negative/null results: article has null or negative findings
- **New finding**: article presents original data, models or hypotheses
- **Novel drug target**: article suggests new targets for drug discovery
- Refutation: article disproves previously published data or hypotheses
- **Technical advance**: article introduces a new practical/theoretical technique, or novel use of an existing technique

The classifications in bold reflect aspects of novelty in research; our expectations are that they are related positively to the novelty scores. For example, we expect that a paper with many assignments of the "new finding" tag from FMs will also have high novelty scores of the indicators considered in this study. The expectations for the tags which are not printed in bold are zero or negative correlations with the novelty scores (e.g., for the confirmation of previously published hypotheses). The "controversial" tag is in italics since it is not entirely clear whether it reflects novelty or not. The tag is basically applied to studies that in the opinion of a FM 'goes against the grain' with regard to current understanding. The finding reported in the paper should be interpreted as 'circumspect', namely one need to carefully think about it before being widely accepted. F1000Prime further assigns the following publication types to the papers: "clinical trial", "systematic review/meta-analysis" and "review/commentary". However, these further assignments are not relevant to the research question of the current study and have not been considered.

---

[1] The definitions of the tags are adopted from https://f1000.com/prime/about/whatis/how

F1000Prime recommendations have been used already in several studies for the correlation with metrics. One such study was interested in the relationship of quantitative (metrics-based) and qualitative (human-based) assessments of research. Waltman and Costas (2014) found "a clear correlation between F1000 recommendations and citations. However, the correlation is relatively weak" (p. 433). A similar result was published by Mohammadi and Thelwall (2013). Bornmann and Leydesdorff (2013) tested the correlation between several bibliometric indicators and F1000Prime recommendations and found that the "percentile in subject area achieves the highest correlation with F1000 ratings" (p. 286). Bornmann (2015) used logistic regression models to examine the convergent validity of the F1000Prime peer review system. His results show that "the proportion of highly cited papers among those selected by the faculty members is significantly higher than expected. In addition, better recommendation scores are also associated with higher performing papers" (p. 2415).

Anon (2005) reports the following additional results which are based on the journal rather than the paper level (see also Jennings, 2006): "papers from high-profile journals tended to be rated more highly by the faculty; there was a tight correlation ($R^2 = 0.93$) between average score and the 2003 impact factor of the journal". Du, Tang, and Wu (2016) differentiated between non-primary research or evidence-based research publications on one side as well as translational research or transformative research publications on the other side. According to their results, the former are more highly cited rather than highly recommended by FMs, and the latter are more highly recommended than highly cited.

### 3.2 Measuring convergent validity

Validity is a key concept in psychometrics; it is the extent to which metrics, indicators, assessments etc. measure what they claim to measure. Different types of validity exist (e.g., face validity, content validity, construct validity or predictive validity). Convergent validity belongs to the construct validity type and is the degree to which two metrics, indicators, assessments etc. of constructs (e.g., productivity, quality, novelty or creativity) which theoretically should be related are in fact (i.e., empirically) related (Panel for Review of Best Practices in Assessment of Research et al., 2012).

Several authors in bibliometrics have argued that the validity of bibliometric indicators should be tested by correlating them with judgements by peers (see e.g., Adams, Loach, & Szomszor, 2016; Kim & Diesner, 2015). Thelwall (2017) explains the background of those tests as follows: "If indicators tend to give scores that agree to a large extent with human judgements then it would be reasonable to replace human judgements with them when a decision is not important enough to justify the time necessary for experts to read the articles in question. Indicators can be useful when the value of an assessment is not great enough to justify the time needed by experts to make human judgements" (p. 4). Kreiman and Maunsell (2011) formulated the following expectations for the test results: "In general, one would like to observe that the metric correlates with expert evaluations across a broad range of individuals with different degrees of productivity". Rowlands (2018) justifies the analysis of convergent validity of indicators based on judgements of peers as follows: "the criteria for convergent validity would not be satisfied in a bibliometric experiment that found little or no correlation between, say, peer review grades and citation measures". Harnad (2008) argues similarly: metrics "need to be jointly tested and validated against what it is that they purport to measure and predict, with each metric weighted according to its contribution to their joint predictive power. The natural criterion against which to validate metrics is expert evaluation by peers" (p. 103).

It is the purpose of the current study to analyze the convergent validity of the indicators introduced to measure the novelty of papers. We generated the scores for two indicators based on the explanations by Uzzi et al. (2013) – revised by Lee et al. (2015) – and Wang et al. (2017) for papers published in the biomedical area. We matched the papers with expert judgements (by FMs) from the F1000Prime database. We were especially interested in their tag assignments to papers which reflect the novelty of reported research (e.g., "article suggests new targets for drug discovery"). We tested whether papers receiving these tags have higher novelty scores than papers without these tags.

### 3.3 Datasets and computations

In order to test the convergent validity of the novelty scores, we calculated the scores for all papers in the F1000Prime dataset that were published between 2013 and 2015. We used three recent publication years to have a sufficiently large dataset for studying the convergent validity. Two novelty scores included in our analyses are based on the combinations of journals in the cited references of the focal papers, as well as the combinations of journals in the cited references of papers that are not among the focal papers. We retrieved these additional data from an in-house database based on the Web of Science (WoS) database from Clarivate Analytics. This database covers papers since 1980, i.e. only cited references published since 1980 are considered in our analyses. Besides the two aforementioned novelty scores based on the focals' cited references, we calculated a simple score measuring novelty based on keyword combinations. The keywords used for this score were also retrieved from our in-house database.

*Novelty score U*

The first novelty indicator denoted as "novelty score U" has been proposed by Uzzi et al. (2013) and revised by Lee et al. (2015). First, at the journal level, they define the commonness of each journal pair (journal *i* and journal *j*) in year *t* as

$$Commonness_{ijt} = \frac{N_{ijt}N_t}{N_{it}N_{jt}} \qquad (1)$$

where $N_{ijt}$ is the number of pairs of cited references where the two references have been published in the journals $i$ and $j$ and co-cited in year $t$, $N_{it}$ is the number of pairs of references where one of the references has been published in journal $i$ and the pair has been co-cited in year $t$, and $N_t$ is the number of all pairs of references that have been co-cited in year $t$. Second, Lee et al. (2015) choose the $10_{th}$ percentile of the series of commonness values for the pairs of journals corresponding to the pairs of cited references of a paper as an indicator of commonness at the paper level. The novelty score is then defined as the negative natural logarithm of this commonness score at the paper level. Lee et al. (2015) provide a detailed explanation of the indicator.

*Novelty score W*

The second indicator denoted as "novelty score W" has been introduced by Wang et al. (2017). For the calculation of this score, all (distinct) pairs of referenced journals of the focal paper are considered, if they fulfil the following conditions: (1) the journal pair has not been co-cited prior to the focal paper; (2) the journal pair is co-cited at least once in the three years following the focal paper's publication year. For each of these new journal pairs, it is measured how easy it is to make this new combination of the two journals by comparing their co-cited journals. If no or only a few other journals have been co-cited with both journals of the new journal pair, this is considered as an indicator that it is not easy to make this new combination (which would be an indicator for a rather novel focal paper).

In contrast, if both journals of the new pair of referenced journals have been co-cited with a third journal many times, this is considered as an indicator that it is easy to make this new combination (which would be an indicator for a less novel focal paper). The ease of combining the two journals of a new journal pair is defined as the cosine similarity between the vectors containing the numbers of co-citations of the two journals with other journals (co-citation profiles). The co-citation profiles are determined based on journal pairs co-cited in the three years prior to the focal paper's publication year. In all calculations, the 50% least cited journals (based on citations over the three years prior to the focal paper's publication year) are excluded. Finally, the novelty score for a focal paper is defined as

$$W = \sum_{new-pair J_i, J_j} (1 - \cos J_{ij}) \qquad (2)$$

where $\cos J_{ij}$ denotes the cosine similarity between the co-citation profiles of journals $J_i$ and $J_j$. The sum is taken over all new pairs of referenced journal pairs as described above.

*Keyword score K*

The two aforementioned metrics to measure novelty are based on novel combinations of cited references. Since Tahamtan and Bornmann (2018) have questioned the use of cited references for measuring novelty and for tracing knowledge flow in general, we additionally included a relatively simple metric for measuring novelty in this study, which is not based on novel cited references, but on keyword combinations (see above). We compared this metric with the novelty scores based on cited references: do we receive other results if we use a score which is not based on cited references, but on keywords?

The keyword score $K_j$ for a given paper $j$ is defined as

$$K_j = \frac{k_{jn}}{k_j} \qquad (3)$$

where $k_{jn}$ is the number of new keywords in a paper $j$ and $k_j$ is the total number of keywords in paper $j$. Therefore, keyword score K is the proportion of new keywords whereby newness is judged against the available keywords in one subject category from the same publication year. If a paper belongs to more than one subject category, K has been calculated multiple times, and we selected the maximum score for the further statistical analysis.

For our final dataset, the novelty scores were matched with the recommendations by FMs of the F1000Prime dataset and the number of citations from our in-house database. The final dataset comprises the novelty scores W and U, the recommendations by FMs and the number of citations received for 23,206 papers. Since the keyword score K is available for a slightly reduced set of 23,041 papers (the missing values are due to lacking keywords), we could not perform all statistical analyses with the final dataset.

**3.4    Statistics**

In this study, we tested the convergent validity of metrics measuring novelty by using tag assignments by FMs. The variable "tag assignments" is a count variable recording how many times a specific tag (e.g., "new finding") has been assigned to a paper by FMs. We used regression models to investigate whether (and to what extent) the assignments depend on the novelty of the papers – as measured by the various novelty scores (see section 2). Regression models for count outcomes are based on the Poisson distribution (Long & Freese, 2014).

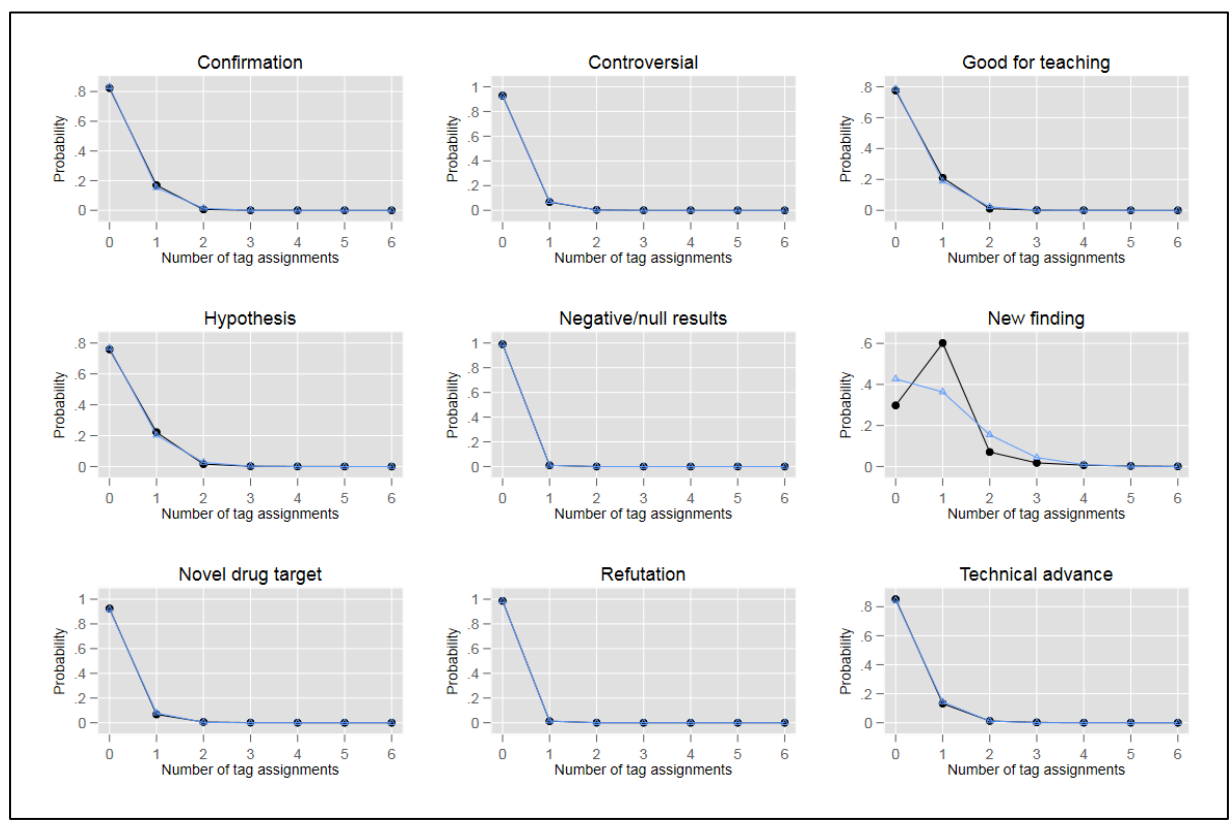

Figure 1. Comparisons of the relative number of tag assignments (black line) with Poisson predictions (blue line)

Figure 1 shows graphs which compare the observed relative frequencies of tag assignments with the predicted probabilities from Poisson regressions (intercept-only models, see below). The figures are based on the numbers of the different tags (e.g., "new finding" or "technical advance") assigned to the papers. The results demonstrate the similarity of the observed and the predicted distributions. The Poisson distribution only underpredicts count 1 in the comparison with the observed "new finding" assignments in Figure 1.

The Poisson regression model (PRM) assumes that the observed count of paper $i$ is drawn from a Poisson distribution with mean $\mu_i$, where $\mu_i$ is dependent on the regressors of the model (i.e., the novelty score in our models). With one independent variable, the PRM is

$$\mu_i = \exp(\beta_0 + \beta_1 x_{i1}) \quad (3)$$

Given our dependent variables (e.g., "new finding" or "good for teaching" assignments), we cannot assume that each paper was at risk of receiving tags for the same amount of time. Since the papers in our dataset are from different years, the papers have different exposure times. The different amounts of time between publication and date of data compilation clearly affect the number of tags. To adjust for exposure time, we additionally included the publication years of the papers in the model. Since "publication year" is a factor variable (i.e., a variable with three categories), we included two variables as further independent variables in the regression models (besides the novelty and keyword scores). The third category of "publication year" is the reference category. The formula of the models changes as follows

$$\text{PRM:}\ \mu_i = \exp(\beta_0 + \beta_1 x_{i1} + \beta_2 x_{i2} + \beta_3 x_{i3}) \qquad (4)$$

In the interpretations of the results of the regression models, we do not focus on the estimates of the regression coefficients. We further calculated – based on these estimates – percentage change coefficients for a standard deviation increase in the independent variable (e.g., novelty score U). The percentage change coefficients simplify the interpretation and comparison of the regression models' results.

Since the novelty and keyword scores as well as citation counts and peers assessments do not follow normal distributions, we decided to calculate regression models with robust standard errors. The robust option in Stata is a post-estimation modification which influences standard error estimations after the point estimation is complete (Angeles, Cronin, Guilkey, Lance, & Sullivan, 2014). The option obtains unbiased standard errors of coefficients in regression analyses; it estimates the variance-covariance matrix of errors in such a way that does not assume normality (Acock, 2018). According to Hilbe (2014), "most researchers use regular model standard errors as the default and employ robust and profile techniques when it is clear that model assumptions have been violated" (p. 99).

Citation counts and assessments by peers are usually used as proxies for the quality of papers (Martin & Irvine, 1983). In this study, we compare the results based on the novelty scores with the results based on citation counts and peers assessments. The variable "FMs recommendations" is the sum of the scores (1, 2 or 3 stars) which the papers receive by FMs. We were interested in how the results of the regression models for the novelty scores differ from those for the quality assessments (using citation counts and peers assessments, respectively, as independent variables instead of novelty and keyword scores). Whereas FM recommendations are quality assessments of papers, citation counts reflect one part of quality: the impact of research within the scholarly community (Martin & Irvine, 1983). Since all F1000Prime tags are used by FMs to point to various factors explaining why the tagged papers are so important for the scientific community, we expect for all tags a positive relationship between the assignments of tags and both FM recommendations and citations, respectively.

Citation counts and FMs recommendations are count variables which can be equally analyzed as the number of tag assignments with PRMs.

# 4    Results

In this study, we empirically investigated the convergent validity of both novelty scores (U and W) by using tag assignments by FMs for comparison. We also examined keyword score K for comparison. Table 1 presents the key figures for the novelty scores, keyword score K, citations, and FMs recommendations. As the results show, both novelty scores have very different values: for novelty score U, the minimum value is a negative value and the maximum is 6.96. In contrast, the minimum value of novelty score W is -152.86 and the maximum score is 303.46. As the mean novelty scores in Table 1 reveal, the novelty of the published research are similar across the publication years.

Table 1. Key figures for both novelty scores, keyword score K, citations, and FMs recommendations (by publication year)

| Statistic | Novelty score U | Novelty score W | Keyword score K | Citations | FMs recommendations |
|---|---|---|---|---|---|
| **2013** | | | | | |
| Mean | 0.69 | 0.37 | 0.17 | 72.59 | 2.23 |
| Median | 0.69 | 0 | 0.13 | 37 | 2 |
| Standard deviation | 0.98 | 4.72 | 0.16 | 159.22 | 1.9 |
| Minimum | -5.47 | -152.86 | 0 | 0 | 1 |
| Maximum | 5 | 303.46 | 1 | 8562 | 35 |
| Number of papers | 9246 | 9246 | 9167 | 9246 | 9246 |
| **2014** | | | | | |
| Mean | 0.68 | 0.34 | 0.17 | 59.63 | 1.96 |
| Median | 0.73 | 0 | 0.13 | 31 | 2 |
| Standard deviation | 0.94 | 1.62 | 0.16 | 100.6 | 1.45 |
| Minimum | -6.27 | -17.14 | 0 | 0 | 1 |
| Maximum | 4.75 | 46.26 | 1 | 2173 | 28 |
| Number of papers | 7395 | 7395 | 7355 | 7395 | 7395 |
| **2015** | | | | | |
| Mean | 0.71 | 0.4 | 0.17 | 47.68 | 1.86 |
| Median | 0.71 | 0 | 0.13 | 24 | 2 |
| Standard deviation | 0.9 | 2.74 | 0.16 | 106.34 | 1.36 |
| Minimum | -5.74 | -0.43 | 0 | 0 | 1 |
| Maximum | 6.96 | 106.13 | 1 | 5419 | 30 |
| Number of papers | 6565 | 6565 | 6519 | 6565 | 6565 |
| **All years** | | | | | |
| Mean | 0.69 | 0.37 | 0.17 | 61.41 | 2.04 |
| Median | 0.71 | 0 | 0.13 | 31 | 2 |
| Standard deviation | 0.95 | 3.44 | 0.16 | 128.95 | 1.63 |
| Minimum | -6.27 | -152.86 | 0 | 0 | 1 |
| Maximum | 6.96 | 303.46 | 1 | 8562 | 35 |
| Number of papers | 23,206 | 23,206 | 23,041 | 23,206 | 23,206 |

For investigating convergent validity, we performed regression models with novelty or keyword scores as independent, and the number of tag assignments by the FMs as dependent variables. Our expectations for the regression analyses were that the number of assignments for the tags "hypothesis" (article presents an interesting hypothesis), "new finding" (article presents original data, models or hypotheses), "novel drug target" (article suggests new targets for drug discovery), and "technical advance" (article introduces a new practical/ theoretical technique) are positively related to novelty and keyword scores. More assignments of these tags which are associated with newness in different forms should appear alongside higher novelty scores. For "confirmation", "good for teaching", "negative/ null results", and "refutation", we expect small or negative correlations with the novelty and keyword scores. We do not have clear expectations for "controversial".

Table 1 additionally includes two variables which we used as independent variables in the regression models besides the novelty and keyword scores: the number of citations and FMs recommendations (both measured since publication until the end of 2018). The FMs recommendations does not refer to the number of recommendations received, but the sum of the scores given by the FMs (see above). The results of the additional regression models are used to compare findings based on novelty and keyword scores with findings based on indicators reflecting quality.

Table 2. Key figures for the dependent and independent variables included in the PRM (which is presented in detail, $n$=23,206)

| Variable | Mean or percent | Standard deviation | Minimum | Maximum |
| --- | --- | --- | --- | --- |
| Dependent variable | | | | |
| Number of "new finding" assignments | 0.85 | 0.77 | 0 | 11 |
| Independent variable | | | | |
| Novelty score U | 0.69 | 0.95 | -6.28 | 6.96 |
| Publication year | | | | |
| 2013 | 40% | | | |
| 2014 | 32% | | 0 | 1 |
| 2015 | 28% | | 0 | 1 |

We calculated several regression models for the different dependent (F1000 FMs' tag assignments) and independent (novelty and keyword scores, FMs recommendations, and citations) variables. Since the number of models is too large to present all of them in detail, we show the results of one model in detail and below focus on the most important (and comparable) results from all models. The model for the detailed analysis refers to "new finding" as dependent and "novelty score U" as independent variables. Table 2 shows the key figures for the dependent and independent variables. The mean value for the "new finding" tag is 0.85 (i.e., the papers received about one "new finding" tag on average); 40% of the papers were published in 2013 and 28% in 2015 (see the variable "publication year").

Table 3. Results of the PRM with the number of "new finding" assignments as dependent variable (which is presented in detail, n=23,206, Pseudo $R^2$=0.33%)

| Independent variable | Coefficient | Robust standard error | Percentage change in expected count for a standard deviation increase in novelty score U (with standard deviation, SD) |
|---|---|---|---|
| Novelty score U | 0.08*** | 0.01 | 7.5 [SD=0.95] |
| Publication year | | | |
| 2014 | -0.08*** | 0.01 | |
| 2015 | -0.15*** | 0.01 | |
| Constant | -0.15*** | 0.01 | |

Note. *** p<0.001

The results of the PRM are presented in Table 3. The estimated coefficient from the PRM for novelty score U is statistically significant and positive. The percentage change data in the table reveals the size of the effect: for a standard deviation increase in novelty score U (i.e., an increase with the value 1), the expected number of assignments of the "new finding" tag increase by 7.5%, holding publication years constant. The standard deviation for novelty score U is approximately one. Thus, the effect of the independent variable is as expected and substantial. It is an advantage of the percentage change data that it is comparable not only across different F1000Prime tags (the result is standardized by using the standard deviation), but also across the models, which are based on keyword scores, FMs recommendations and citations. The standard deviation has the same meaning in the independent variables.

Table 4 shows the results of all regression models which have been performed in this study. The models are based on different tags as dependent variables and novelty and keyword scores, citations, and FMs recommendations as independent variables (besides the publication years). The table also includes explanations of the F1000Prime tags as well as our expectations for the novelty scores' results. These expectations refer to the direction of the relationship between the novelty scores and tag assignments (see section 3.1).

Table 4. Results of PRMs (percentage change data) with the number of assignments of different tags as dependent variables and novelty and keyword scores, citation counts, or FMs recommendations as independent variables (in addition to publication years).

| Tag | Explanation | Expected results for the novelty score | Tag | Novelty score U | Novelty score W | | Citations | FMs recommendations | Keyword score K | |
|---|---|---|---|---|---|---|---|---|---|---|
| | | | Percent assigned | Percentage change | Percentage change | Percentage change (W as categorical variable) | Percentage change | Percentage change | Percent assigned | Percentage change |
| Confirmation | Validates previously published data or hypotheses | Negative | 17.61 | -13.45*** [0.58%] | -0.71 [0.14%] | -1.25 [0.14%] | 1.19 [0.14%] | 14.92*** [0.77%] | 17.61 | -3.39* [0.16%] |
| Good for teaching | Key article in field and/or well written | Negative | 22.39 | -2.06 [0.02%] | 3.26*** [0.09%] | 10.00*** [0.23%] | 5.66*** [0.28%] | 27.10*** [3.92%] | 22.33 | 3.71*** [0.04%] |
| Negative/Null results | Article has null or negative findings | Negative | 0.97 | -19.86*** [1.06%] | -4.84** [0.53%] | -6.66 [0.55%] | -20.17 [0.66%] | 11.08* [0.65%] | 0.97 | -9.14 [0.59%] |
| Refutation | Disproves previously published data or hypotheses | Negative | 1.33 | -1.14 [0.14%] | 1.81 [0.15%] | -1.09 [0.14%] | 4.84*** [0.21%] | 28.55*** [2.39%] | 1.34 | 5.73 [0.17%] |
| Controversial | Challenges established dogmas | Negative/ Positive | 7.13 | -10.35*** [0.38%] | 2.36* [0.22%] | 5.25* [0.24%] | 4.65*** [0.29%] | 27.28*** [3.01%] | 7.08 | -0.91 [0.21%] |
| Hypothesis | Article presents an interesting hypothesis | Positive | 24.09 | 17.71*** [0.58%] | 1.6*** [0.09%] | 2.91* [0.09%] | 4.67*** [0.21%] | 28.45*** [4.95%] | 24.13 | -1.32 [0.08%] |
| New finding | Presents original data, models or hypotheses | Positive | 70.21 | 7.47*** [0.33%] | -3.91*** [0.20%] | -8.47*** [0.40%] | 5.51*** [0.61%] | 26.66*** [7.27%] | 70.38 | -6.55*** [0.31%] |
| Novel drug Target | Suggests new targets for drug discovery | Positive | 7.55 | 58.78*** [2.84%] | -4.10*** [0.20%] | -6.27* [0.23%] | 6.81*** [0.62%] | 32.84*** [5.95%] | 7.59 | -12.15*** [0.39%] |
| Technical advance | Introduces a new practical/ theoretical technique | Positive | 14.82 | 28.78*** [1.02%] | -0.13 [0.03%] | -2.61 [0.04%] | 7.38*** [0.83%] | 30.78*** [5.48%] | 14.80 | -7.83*** [0.14%] |
| Number of papers | | | 23,206 | 23,206 | 23,206 | 23,206 | 23,206 | 23,206 | 23,041 | 23,041 |

Notes. *** p<0.001, ** p<0.01, * p<0.05. Pseudo $R^2$ is reported in each cell in square brackets.

Let us start the interpretation of the results in Table 4 with the findings for citations and FMs recommendations. Both variables are expected to reflect the "quality" dimension of published research and we predicted a positive relationship with all tag assignments. As the results in the table reveal, the expectations are met for all tags, but with one exception: papers with negative or null results are expected to have fewer citations than papers with positive results. However, this is something which has been reported on citations previously. For example, Jannot, Agoritsas, Gayet-Ageron, and Perneger (2013) found a citation bias favoring significant results in medical research.

The percentage change data in Table 4 is larger for FMs recommendations than for citations. Since the tags are from the same experts as the recommendations, higher percentage change values can be expected for the experts than for the citations.

The results in Table 4 for novelty score U completely match the expectations: all signs of the percentage change data are in the expected directions. Thus, the indicator "novelty score U" seems to have convergent validity with the assessments by peers. As the results further reveal, the indicator seems to be especially suitable for identifying novelty in targets for drug discovery. The percentage change value is very high (58.78%). However, larger percentage change values are also visible for other tags reflecting newness: "hypothesis" (17.71%) – although statistically not significant – and "technical advance" (28.78%). The percentage change value for "new finding" is comparatively low (7.47%). This might be the case because the tag has been very frequently assigned to papers by the experts: 70.21% of the papers received the tag "new finding".

The findings for novelty score W are not as favorable as those for novelty score U. Only the results for three tags are as expected: "confirmation", "negative/null results", and "hypothesis". Furthermore, the percentage change data for novelty score W turned out to be generally lower than those for novelty score U. It seems that novelty score W is not as convergently valid as novelty score U. We calculated additional regression models for novelty score W for comparison, since Wang et al. (2017) generated and used a categorical novelty variable in their study as follows:

(1) non-novel papers (if a paper has no new journal combinations),

(2) moderately novel papers (if a paper makes at least one new combination but has a novelty score lower than the top 1% in our dataset), and

(3) highly novel paper (if a paper has a novelty score among the top 1% in our dataset).

Wang et al. (2017) calculated this variable in a slightly other way, since they used all papers of the corresponding subject categories to identify the top 1%. We did not deem it as necessary to use all papers from corresponding subject categories for determining the top 1%, since all papers in our dataset are from the biomedical area.

As the findings in Table 4 for the categorical novelty variable W reveal, they also do not agree with the expectations for many tags. The results only change slightly in comparison to the novelty score W variable.

For comparison with the novelty scores based on cited references, we calculated keyword scores K. As the results in Table 4 reveal, keyword scores K perform similar unsatisfactorily as novelty score W: the results for only two tags are in the expected direction.

Table 5. Results of probit regressions models (fully standardized coefficients) with the assignments of different tags as dependent variables and novelty scores as independent variables (besides publication years)

| Tag | Explanation | Expected results for the novelty score | Fully standardized coefficients | | |
| --- | --- | --- | --- | --- | --- |
| | | | Novelty score U | Novelty score W (W as categorical variable) | Keyword score K |
| Confirmation | Validates previously published data or hypotheses | Negative | -0.10*** [0.70%] | -0.01 [0.17%] | -0.02 [0.17%] |
| Good for Teaching | Key article in field and/or well written | Negative | -0.03** [0.05%] | 0.07*** [0.27%] | 0.04*** [0.09%] |

| Negative/Null results | Article has null or negative findings | Negative | -0.09*** [1.13%] | -0.02 [0.59%] | -0.04 [0.63%] |
|---|---|---|---|---|---|
| Refutation | Disproves previously published data or hypotheses | Negative | -0.01 [0.14%] | -0.01 [0.15%] | 0.01 [0.15%] |
| Controversial | Challenges established dogma | Negative/ Positive | -0.06*** [0.47%] | 0.02 [0.25%] | -0.01 [0.23%] |
| Hypothesis | Article presents an interesting hypothesis | Positive | 0.12*** [0.73%] | 0.03*** [0.08%] | 0.01 [0.05%] |
| New Finding | Presents original data, models or hypotheses | Positive | 0.12*** [0.79%] | -0.12*** [0.76%] | -0.04*** [0.16%] |
| Novel Drug Target | Suggests new targets for drug discovery | Positive | 0.27*** [3.26%] | -0.03* [0.15%] | -0.06*** [0.29%] |
| Technical advance | Introduces a new practical/ theoretical technique | Positive | 0.16*** [1.22%] | 0.00 [0.01%] | -0.03** [0.04%] |
| Number of papers | | | 23,206 | 23,206 | 23,041 |

Note. *** p<0.001, ** p<0.01, * p<0.05. Pseudo $R^2$ is reported in each cell in square brackets.

In a further step of the empirical analyses, we investigated the robustness of our results. We additionally performed probit regression models which can be used to model binary outcome variables. In these models, the inverse standard normal distribution of probability is modeled as a linear combination of the independent variables (Long & Freese, 2014). In the previous PRMs, we used the sum of the FMs scores as dependent variables. For the robustness checks, the data have been dichotomized showing whether a paper received a certain tag or not. These binary variables can be analyzed with probit regression models. In the regression models based on novelty score W, we included W as categorical variable, since some models based on W did not converge.

The results of the models for novelty scores U and W as well as keyword score K are reported in Table 5. The fully standardized coefficients can be interpreted as follows (here for the novelty score U and "new finding" variables): for every standard deviation increase in novelty score U, a paper's propensity to receive the "new finding" tag is expected to increase by 0.12% standard deviations, holding the publication year constant. As the results of all regression models in the table reveal, the conclusions which we formulated previously based on the PRMs do not change substantially with the new results. The further results mainly confirm the robustness of the PRMs.

In the final step of the empirical analyses, we compared novelty and keyword scores for two further groups of papers to test the robustness of our main findings reported previously. The results in Table 4 reveal that 70.21% of the F1000Prime papers have the "new finding" tag. We can infer from this result that "newness" is a typical sign of papers included in the F1000Prime database. We compared therefore the novelty and keywords scores of two groups: (1) included in F1000Prime and (2) not included in F1000Prime. The group of papers included in F1000Prime consist of all papers in the F1000Prime database of document type article or review that have been published between 2013 and 2015. This comprises 22,832 papers. The group of papers not included in F1000Prime was retrieved by drawing a random sample of the same size, stratified by publication years and document types (considering the distribution in the group of papers included in F1000Prime). For drawing the sample, we could not consider all papers published in these years, since the FMs only select papers from biomedicine and related areas. Waltman and Costas (2014) solved this problem by using the following procedure in their study on F1000Prime: they identified all subject categories of the papers included in F1000Prime and selected all papers which are in at least one subject category previously identified. We followed this procedure in our study.

Table 6. Key figures for the dependent (included in F1000Prime) and independent variables (novelty and keyword scores) included in three logistic regression models

| Variables | Mean or percent | Standard deviation | Minimum | Maximum |
|---|---|---|---|---|
| 2013 (n=17,887) | | | | |
| Included in F1000Prime | 0.51% | 0.50 | 0 | 1 |
| U | 0.25 | 1.38 | -5.71 | 11.41 |
| W | 0.42 | 3.81 | -152.86 | 303.46 |
| K | 0.21 | 0.19 | 0 | 1 |
| 2014 (n=14,345) | | | | |

| Included in F1000Prime | 0.51% | 0.50 | 0 | 1 |
| U | 0.35 | 0.74 | -6.28 | 4.75 |
| W | 0.39 | 1.97 | -33.62 | 107.18 |
| K | 0.21 | 0.18 | 0 | 1 |
| 2015 (n=12,697) | | | | |
| Included in F1000Prime | 0.51% | 0.50 | 0 | 1 |
| U | 0.37 | 0.73 | -5.74 | 6.96 |
| W | 0.46 | 2.63 | -0.43 | 106.13 |
| K | 0.21 | 0.18 | 0 | 1 |

We performed logistic regression models (with robust standard errors) to determine the relationship between novelty and keyword scores and the inclusion of papers in F1000Prime. We performed logistic instead of probit regression models (see above), since the probit regression models did not converge in all cases. In the logistic regression models, we used the binary variably "included in F1000Prime" as dependent (see above) and the novelty and keyword scores as independent variables. We included all novelty and keyword scores in one model (and calculated three models for the publication years 2013, 2014, and 2015). The consideration of the variables in separate models show that the results are very similar to those of the full model (the spearman rank-order correlations between the novelty and keyword scores are between -0.05 and 0.1). The key figures for the variables included in the models are shown in Table 6.

Table 7. Results of the logistic regression models with the binary variable "included in F1000Prime" as dependent variable

| Variable | Coefficient | Robust standard error | Percentage change in odds for a standard deviation increase in the score |
|---|---|---|---|
| 2013 (n=17,887 Pseudo $R^2$=11.65%) | | | |
| Novelty score U | 0.59*** | 0.02 | 124.2 |
| Novelty score W | 0.00 | 0.00 | 0.7 |
| Keyword score K | -2.47*** | 0.10 | -37.2 |
| Constant | 0.38*** | 0.02 | |
| 2014 (n=14,345, Pseudo $R^2$=24.54%) | | | |
| Novelty score U | 2.39*** | 0.11 | 489.5 |
| Novelty score W | -0.10 | 0.02 | -18.3 |
| Keyword score K | -2.54*** | 0.11 | -37.3 |
| Constant | -0.12** | 0.04 | |
| 2015 (n=12,697, Pseudo $R^2$=28.51%) | | | |
| Novelty score U | 2.84*** | 0.14 | 689.3 |
| Novelty score W | -0.06*** | 0.02 | -15.1 |
| Keyword score K | -2.51*** | 0.12 | -37.1 |
| Constant | -0.23*** | 0.04 | |

Notes. ** p<0.01, *** p<0.001

The results of the three regression models (for 2013, 2014, and 2015) are shown in Table 7. Besides the coefficients, percentage changes in expected counts for being included in F1000Prime are shown. The results for the novelty scores reinforce the previous results. A standard deviation increase in the novelty scores increases the odds of being included in F1000Prime by more than 100% in case of novelty score U. All coefficients and percent change values for keyword score K and two of three coefficients and percent change values for novelty score W in Table 7 are negative (2014 and 2015): a standard deviation increase in the scores decreases the odds of being included in F1000Prime. Only the 2013 coefficient and percent change value for keyword score W are positive.

## 5 Discussion

Creativity is the essence in which breakthrough-class research emerges. Research without creative inspiration might become "normal science" – according to the notation introduced by Kuhn (1962) – but will not be exceptionally enough to be "revolutionary". The novel and unusual combination of existing knowledge from the archives has been suggested as act of creativity. Citation data are especially suited to measuring the use of novel (or typical) combinations for certain pieces of research, since it is usual in science to cite the shoulders on which research stands (Merton, 1965): according to the ethos of science, all publications which have influenced the reported research should be cited in a manuscript (Merton, 1973). Thus, the novelty of research might be reflected in the typical – as many others do – or novel – as only a few others do – use of cited references. Lee et al. (2015) – based on Uzzi et al. (2013) – and Wang et al. (2017) proposed scores based on cited references (cited journal) data which can be used to measure the novelty of papers. Whereas Uzzi et al. (2013) identifies <u>atypical</u> combinations of cited journal pairs, Wang et al. (2017) focus on <u>novel</u> combinations. According to Lee et al. (2015), "it is important to find indicators that allow us to unpack the concepts of novelty and impact, in order to better understand the drivers of creativity" (p. 685).

In previous studies, the success of the scores to measure novelty has been investigated by citation analyses (see the overview of Zeng et al., 2017). The results of Uzzi et al. (2013) including "17.9 million papers across all scientific fields suggest […] that the highest-impact science draws on primarily highly conventional combinations of prior work, with an intrusion of combinations unlikely to have been joined together before. These patterns suggest that novelty and conventionality are not factors in opposition; rather, papers that mix high tail novelty with high median conventionality have nearly twice the propensity to be unusually highly cited" (p. 471). The results by Wang et al. (2017) reveal that papers with high novelty scores tend to be less cited after their appearance, but are more highly cited after several years than papers with low scores. Furthermore, novel research is "not only more likely to become a big hit itself but also more likely to stimulate follow-on research which generates major impact" (Wang et al., 2017, p. 1420). Mairesse and Pezzoni (2018) used a novelty score based on cited references data to study the impact of "standard" and "novel" papers published by French scientists. They also found – in agreement with Wang et al. (2017) – that novelty is reflected in citations only in the long run.

Although previous research has used novelty scores in various empirical analyses, only one study has been published up to now – to the best of our knowledge – which investigated whether novelty (or creativity) is really rooted in cited references (see section 1). However, the authors – Tahamtan and Bornmann (2018) – undertook a qualitative study based on interview data. The current study tested empirically (quantitatively) the convergent validity of novelty scores: do these scores measure what they propose to measure? The investigation of validity is an important step in the development of new indicators (Thelwall, 2017); only validated indicators should be used in the practical application. Using novelty assessments by FMs of F1000Prime for comparison, we tested in this study the convergent validity of three novelty scores (U, W, and K). Peers' assessments at F1000Prime do not only refer to the quality of biomedical papers, but also their characteristics (by assigning certain tags to the papers): for example, are the presented findings or formulated hypotheses novel? Does the paper present a good overview of research on a certain topic explaining why it can be used for teaching activities? We used these tags to investigate the convergent validity of the novelty scores.

Our study reveals different results for the novelty scores: the results for novelty score U are (mostly) in agreement with our expectations concerning the results for the different tags. We found, for instance, that for a standard deviation increase in novelty score U, the expected number of assignments of the "new finding" tag increases by 7.47% (the result is statistically significant). The results further show that this indicator seems to be especially suited to identifying papers suggesting new targets for drug discovery. The corresponding percentage change information from the regression model turned out to be very high (58.78%). The results for novelty score W, however, do not reflect convergent validity with the assessments by FMs: only the results for some tags are in agreement with the expectations. Thus, we propose – based on our results – the use of novelty score U for measuring novelty quantitatively, but question the use of novelty score W.

Measuring the convergent validity of newly proposed novelty scores is very much necessary, but the approach used in this study is not without its limitations. (1) Since F1000Prime data focuses on the biomedical field, our results are only valid for this area. Although it would be interesting to also have results for other areas, similar datasets with peer assessments for many papers are not available (to the best of our knowledge). (2) Since our results refer to only two novelty scores based on cited references (U and W), it would be interesting to have similar investigations of the other scores which have been proposed previously. Although most of the other scores are not based on cited references data (but on keywords, MeSh terms etc.), their convergent validity can be tested also with F1000Prime data. With keyword score K, we considered one of these metrics in this study for comparison; however, it is a very basic metric, and more advanced metrics based on keywords, MeSh terms etc. might be considered in future studies. (3) The results of Bornmann, Tekles, and Leydesdorff (2019) suggest that FMs might be influenced by the citations to the paper when they make their judgements. Hence, to the extent that citations influence the assignment of F1000 tags reflecting aspects of novelty in research, there would be measurement errors in these assignments as measures purely of novelty (rather it would be measuring some combination of novelty and impact).

# Acknowledgements

The bibliometric data used in this paper are from an in-house database developed and maintained in collaboration with the Max Planck Digital Library (MPDL, Munich) and derived from the Science Citation Index Expanded (SCI-E), Social Sciences Citation Index (SSCI), Arts and Humanities Citation Index (AHCI) prepared by Clarivate Analytics, formerly the IP & Science business of Thomson Reuters (Philadelphia, Pennsylvania, USA). We acknowledge the National Natural Science Foundation of China Grant 71673131 for financial support. We would like to thank Tom Des Forges and Ros Dignon from F1000 for providing us with the F1000Prime dataset. We thank Adam Y. Ye for helpful discussions on how to process the data, and Jian Wang for his support in calculating novelty score W.